\documentstyle[twocolumn,aps,epsfig]{revtex}

\begin{document}

% \draft command makes pacs numbers print
\draft
\title{Experimental feasibility of measuring the gravitational redshift of
light using dispersion in optical fibers}
% repeat the \author\address pair as needed
\author{S. Manly, E. Page}
\address{Department of Physics and Astronomy}
\address{University of Rochester}
\date{\today}
\maketitle
\begin{abstract}
% insert abstract here

This paper describes a new class of experiments that use
dispersion in optical fibers to convert the gravitational
frequency shift of light into a measurable phase shift or time
delay. Two conceptual models are explored. In the first model,
long counter-propagating pulses are used in a vertical fiber optic
Sagnac interferometer. The second model uses optical solitons in
vertically separated fiber optic storage rings. We discuss the
feasibility of using such an instrument to make a high precision
measurement of the gravitational frequency shift of light.

\end{abstract}
% insert suggested PACS numbers in braces on next line

% body of paper here

\section{INTRODUCTION}
\label{sec-intro}

Einstein's Equivalence Principle (EEP) is the pillar underlying
all metric theories of gravity, including the theory of general
relativity. This principle states that effects of a uniform
gravitational field are identical to the effects of uniform
acceleration of a coordinate system. EEP can be decomposed into
three components: the weak equivalence principle, local Lorentz
invariance, and local position invariance (LPI). Precision
experiments testing each of these components have yielded results
consistent with EEP and the theory of general relativity
\cite{will}\cite{littrell}. However, it is interesting to note
that current solar and atmospheric neutrino data are consistent
with a small violation of EEP \cite{neutrinos}. Given the
fundamental nature of EEP, higher precision tests are desirable.
Such experiments might rule out some metric theories of gravity,
uncover a deviation from our expectation that would require a
modification of our current view of gravity, and place constraints
on new macroscopic forces (which might be perverse enough to
respect the weak equivalence principle, but not LPI)\cite{jaffe}.

Local position invariance requires the results of local
nongravitational experiments to be independent of position and
time. Limits on violations of LPI have been set by searching for
variation in the `constants' of nature and by gravitational
redshift experiments \cite{will}. To date, all experimental
results are consistent with LPI.

In a uniform gravitational field with strength g, the frequency
shift of light traveling a distance z along the direction of
maximum field gradient is given by
\begin{equation}
\nu '\approx \nu \left\{ 1\pm \frac{gz}{c^{2} } \right\}
\end{equation}
to first order, where the negative sign is used for light
traveling against the gravitational field and the positive sign is
used for light traveling with the field. Experiments that measure
the gravitational frequency shift of light report results as a
limit on the deviation observed from the equation above. The
deviation, ${\alpha}$, is defined by
\begin{equation}
\frac{\Delta \nu }{\nu } =(1+\alpha )\frac{\Delta U}{c^{2} },
\end{equation}
where ${\Delta}$U is the change in gravitational potential (gz in
the example above) encountered by the light. Currently, the best
limit on ${\alpha}$ comes from a 1976 experiment that compared the
frequency of a precision hydrogen maser frequency standard on the
surface of the earth with one launched by a Scout Rocket to a
height of approximately 10,000 km. This experiment yielded a
result of ${\alpha}$\texttt{<}2x10$^{-4}$ \cite{vessot}.

\section{A new class of experiments}
\label{sec-newclass}

Continuous wave light will travel a vertical distance h in an
optical fiber with index of refraction n in a time
\begin{equation}
t=\frac{h}{c} n
\end{equation}
The frequency dependence of this flight time is given by
\begin{equation}
dt=\frac{h}{c} \left( \frac{dn}{d\nu } \right) d\nu.
\end{equation}
Assuming this to take place near the earth's surface (let g be
constant), the change in the gravitational potential over a
distance dh induces a frequency change
\begin{equation}
d\nu =\frac{g\nu }{c^{2} } dh.
\end{equation}
Over a height h, the flight time of the light will differ from
that given by equation (3) by an amount
\begin{equation}
\delta t=\frac{g\nu h^{2} }{2c^{3} } \left( \frac{dn}{d\nu }
\right),
\end{equation}

Assuming a sinusoidal form for the light waves traveling down the
fiber optic cable of
\begin{equation}
e^{i(\beta z - \omega t)},
\end{equation}
where
\begin{equation}
\beta = \frac{n \omega}{c}.
\end{equation}
and
\begin{equation}
\omega = 2\pi \nu.
\end{equation}
Defining the phase as
\begin{equation}
\phi = \beta (\omega) z - \omega t,
\end{equation}
the phase variation becomes
\begin{equation}
\delta \phi = (\delta \beta) z + \beta \delta z - (\delta \omega)
t - \omega \delta t.
\end{equation}
Specializing to our case, z become the height difference, h, and
since this is a constant in our proposal, the variation of h is
zero and the second term in the above equation disappears. Now the
propagation parameter, $\beta$ can be expanded, to first order as
\begin{equation}
\beta(\omega) = {\beta_0}({\omega}_0) +
{\beta}_1({\omega}_0)\Delta \omega + .....
\end{equation}
Keeping the first order term in the variation and substituting
this and our previously derived equations for the t's into the
phase equation above, we get
\begin{equation}
\delta \phi = {\beta}_1 \delta \omega h - \frac{nh}{c} \delta
\omega - \omega \frac{g \omega h^2}{2c^3} \frac{dn}{d \omega}
\end{equation}
From the definitions above
\begin{equation}
{\beta}_1 = \frac{d\beta}{d\omega} = \frac{\omega}{c}
\frac{dn}{d\omega} + \frac{n}{c}.
\end{equation}
Using this in the above result for $\delta$$\phi$ gives
\begin{equation}
\delta \phi = \frac{g\omega}{c^2}h(\frac{\omega}{c}
\frac{dn}{d\omega}h + \frac{nh}{c} - \frac{nh}{c}) - w \frac{g
\omega h^2}{2c^3}\frac{dn}{d\omega}
\end{equation}
and
\begin{equation}
\delta \phi =\frac{g {\omega}^2 h^2}{2c^3}\frac{dn}{d\omega}.
\end{equation}

\begin{figure}[b]
\centerline{ \epsfig{file=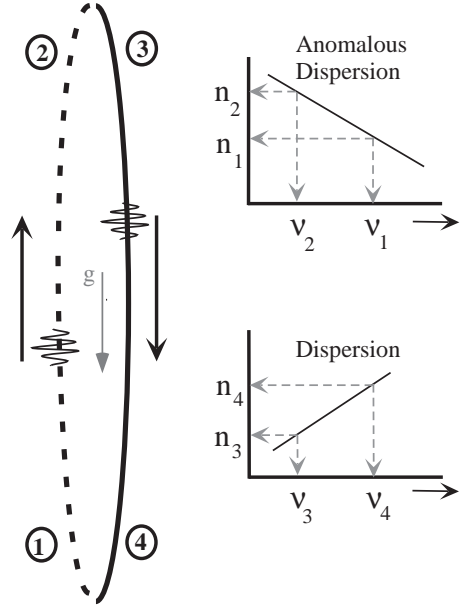, width=8cm} }
   \caption{Diagram illustrating the interplay of the gravitational
   frequency shift of light and dispersion.}
   \label{dispersion_scheme}
\end{figure}

In principle, this phase shift can be measured relative to light
traveling through a fiber with different dispersion to determine
or place a limit on ${\alpha}$ in equation (2). Alternatively, the
information could be used to make a measurement of g. In fact, a
net phase shift can be preserved while the light is returned to
the original height by making the return path out of fiber with a
different dispersion, as shown in Figure~\ref{dispersion_scheme}.
Unfortunately, the phase shift expected per unit length of fiber
is small for reasonable values of dispersion, making it necessary
to have a relatively long path length to produce a measurable
effect. The experimental difficulty is to maintain sufficient
light at the end of the path to make a measurement while
preserving the phase shift or timing well enough to make the
measurement meaningful.

We have considered two experimental models in evaluating the
feasibility of a high precision measurement of the gravitational
redshift of light using this effect. First we considered using
long counter-propagating pulses in a vertical fiber optic Sagnac
interferometer. The second model uses optical solitons propagating
in two vertically separated fiber optic storage rings. In spite of
being less conventional, the latter approach presents fewer
technical challenges and is more likely to yield a robust
measurement.  Since we believe the second of these models is more
experimentally feasible, we present it first.

\section{Experimental concept: vertically separated optical soliton
storage rings}
\label{sec-expt-soliton}

A coherent beam of light pulses is split into two streams, each of
which is transformed into a bitstream of solitons\cite{soliton}.
The streams are injected and stored for a period in separate fiber
optic storage loops \cite{stor_loops} sitting at different
gravitational potentials. The group velocity of the solitons in
the two storage rings will differ due to the interplay of
dispersion with the gravitationally-induced change in the carrier
frequency of the solitons. Unfortunately, the substantial timing
jitter that normally accompanies solitons in storage rings makes
it impractical to measure directly the relative quantum phase
difference between the two beams. Rather, a relative net timing
difference between the pulses in the two beams will develop over
time, which is proportional to the gravitational frequency shift
of the light. This timing difference is used to measure, or place
a limit, on ${\alpha}$.

\subsection{Conceptual details} \label{sec-conceptual-details}

Consider an optical pulse that is split into two coherent pulses
with the same intensity profile. Let one travel horizontally for a
distance h and the other travel vertically a distance h. The group
velocity dispersion parameter ${\beta}$$_{2}$ for each pulse is
given by
\begin{equation}
\beta _{2} =-\frac{1}{2\pi } \frac{dv_{g} }{d\nu } ,
\end{equation}
where v$_{g}$ is the group velocity of the carrier frequency of
the pulse \cite{agrawal}. Assuming the fiber to be identical for
each path, there is a gravitationally-induced group velocity
difference,
\begin{equation}
\left| \Delta v_{g} \right| =\Delta \nu \left( \frac{dv_{g} }{d\nu
} \right) =2\pi \beta _{2} \frac{gh\nu }{c^{2} } ,
\end{equation}
between the pulses as they reach the end of their respective
paths, where g is assumed constant over the height difference. (In
practice, the variation in g must be measured and taken into
account.) At this point, if the pulse streams were each allowed to
travel a horizontal distance d along identical fibers and then
recombined (in such a manner that the path lengths were
identical), there would be a relative pulse timing difference
between the pulses \cite{anadan}. Unfortunately, the effect is
quite small for reasonable input parameters and practical
distances.

\begin{figure}[b]
\centerline{ \epsfig{file=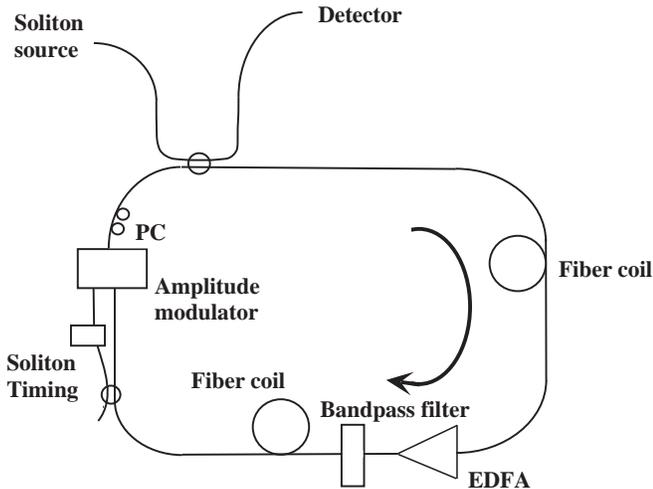,width=9cm} }
   \caption {The basic components of a fiber optic storage loop.}
   \label{basic components}
\end{figure}

One way of increasing the size of the effect is to increase the
distance d. In this model, that is done by using optical soliton
pulses and injecting each stream of pulses into `identical' fiber
optic storage loops, where amplification is provided by erbium
doped fiber amplifiers (EDFA's). The soliton character prevents
dispersion from destroying the pulses, while amplified spontaneous
emission noise is controlled using amplitude modulation, and pulse
timing jitter is reduced with frequency filtering. This model is
schematically illustrated in Figure~\ref{basic components}. This
and similar techniques for pulse memory and storage and long-haul
transmission have been studied extensively in the last decade
\cite{stor_loops}. A scheme similar to what we are proposing was
used in 1993 to store a 10 GHz bit pattern for distances up to
180x10$^{6}$ km (storage times of 15 minutes), as described in
reference \cite{nazakawa}. The authors of that work conclude,
``error free transmission over unlimited distances is possible
with this technique.''

The goal of the telecommunications and pulse storage research on
optical storage loops is to preserve the information, i.e., a high
rate bit pattern, over time and distance. Our goal is somewhat
broader, as we want to preserve the bit pattern \textit{as well
as} a small relative overall group velocity difference between the
bits in two separate streams. If preserved over time, this
relative group velocity difference will be translated into a
measurable pulse timing difference. There are sources of random
pulse timing jitter, as we will discuss below. However, these
noise sources will cause the solitons to random walk in relative
timing, while the gravitationally-induced relative velocity
difference should provide a small, but inexorable, bias in one
direction. There are other effects (as discussed below) which can
cause a unidirectional shift in the group velocity of the
solitons. In this model, the gravitationally-induced effect is
pulled out from the background through differential measurements
of the relative pulse timing between configurations with loops at
different heights.

Figure~\ref{basic components} shows the basic components of a
soliton storage ring experiment. Soliton pulses are used because
these solutions to the nonlinear Schroedinger equation balance
dispersion against the non-linear Kerr effect, and yield a pulse
that is largely immune to dispersive effects in the fiber. Light
loss due to scattering is replaced with continuous or lumped
amplification by EDFAs. Synchronous amplitude modulation limits
the growth of amplified spontaneous noise that would otherwise
overwhelm the signal, and a frequency filter is used in concert
with the modulator to stabilize the pulse from amplitude and width
fluctuations.

\subsection{Noise sources} \label{sec-noise-sources}

There are numerous things that can effect the pulse timing in such
an optical storage loop.
\begin{itemize}
\item \textbf{\textit{Gordon-Haus Jitter}} \\
All optical amplifiers introduce noise into the system by
amplified spontaneous emission (ASE). In non-linear fiber optics,
this ASE noise is also called Gordon-Haus Jitter. When an optical
amplifier introduces ASE, the pulse undergoes small amplitude and
phase fluctuations at the amplifier output, which lead to a change
in the carrier frequency, and therefore the group velocity of the
pulse. Over a span of fiber optic cable, this leads to a random
jitter in the timing of the soliton pulse. The effect of this
jitter on soliton stream transmission has been quantified as
\cite{agrawal} \cite{gordon_haus}
\begin{equation}
BL_{T} <\left[ \frac{\pi f_{j} ^{2} }{q_{0} \lambda h\alpha \gamma
D} \right] ^{1/3} ,
\end{equation}
where B is the bit rate, L$_{T}$ is the length traveled, f$_{j}$
is the fraction of the bit slot that the error should be smaller
than, q$_{0}$ is the initial soliton separation in normalized
units, ${\lambda}$ is the wavelength of light, ${\alpha}$ is the
attenuation coefficient, ${\gamma}$ is the nonlinearity
coefficient, and D is the dispersion coefficient.

Agrawal gives an illustrative example in reference \cite{agrawal}.
A system using 1550 nm light with q$_{0}$ = 10, ${\alpha}$ = 0.2
dB/km, ${\gamma}$ = 10 W$^{-1}$/km, and D = 2 ps/(nm-km), and
assuming f$_{j}$= 0.2 for tolerable jitter. leads to
\begin{equation}
BL_{T} <34,000(Gb/s)-km.
\end{equation}
A 10 Gbit/s system is limited to a transmission length of 3400 km
in this example. This length is small compared to the soliton
transmission lengths required by our experiment.

Fortunately, there are a number of ways to overcome this
bit-length limitation \cite{stor_loops} \cite{agrawal}
\cite{nazakawa} \cite{haus_mecozzi}, including the use of optical
filters and external modulators. Filters reduce the overall noise
level by eliminating the ASE-noise outside the filter bandwidth
and synchronous modulation forces the soliton to realign itself
within the bit slot. Both techniques reduce the jitter, and when
used together properly, can lead to an unlimited propagation
distance.
\item \textbf{\textit{Raman Self Frequency Shift (RSFS)}} \\
As the soliton pulse travels down a fiber optic cable, the photons
scatter off of atoms in the cable. In the process, some of the
photon energy can be lost to atomic vibrational modes, yielding an
outgoing photon frequency that is lower than that of the incoming
photon. This effect, known as the Raman effect, causes a
continuous flow of energy from the high frequency part of the
soliton pulse to the low frequency part of the pulse. Thus, there
is a downshift in the carrier frequency of the soliton as it moves
along the cable, which effects the group velocity of the soliton
and could give a relative timing difference in our experiment if
the effect is asymmetric between the two rings \cite{agrawal}
\cite{gordon}\cite{mitschke}.

It is well known that a light pulse and its frequency range are
related to each other via the Fourier transform. Therefore, the
shorter the soliton pulse, the wider its frequency range. Solitons
of $\sim$1 ps or less have a wide enough frequency range that the
frequency shift of the soliton due to the Raman effect is very
noticeable. Models have shown \cite{gordon} that the redshift
associated with the soliton scales T$_{0}$$^{-4}$, where T$_{0}$
is the soliton pulse width. So, the noticeable effect drops off
quickly for pulse widths above 1 ps. This is an important design
consideration in our experiment.
\item \textbf{\textit{Soliton Interactions}} \\
The same non-linearity that causes solitons to be stable also
causes an interaction force between solitons. The interaction
force between solitons depends on the separation of the solitons,
their relative amplitude, and their relative phase \cite{agrawal}.
Only in the case of two solitons with the same relative phase is
the soliton-soliton interaction attractive. In all other cases,
where the solitons have a different relative phase, the
interaction force is repulsive. In the specific case when the
relative phase ${\theta}$ = 0 and the relative amplitude r = 1,
the relationship between soliton separation and distance traveled
down a fiber optic cable is given by
\begin{equation}
\exp \lbrack 2(q-q_{0} )]=\frac{1}{2} \lbrack 1+\cos (4\xi
e^{-q_{0} } )],
\end{equation}
where q$_{0}$ is the initial soliton separation, and q is the
soliton separation at a distance ${\xi}$ along the fiber. From
this equation it can be seen that the solitons periodically
combine and then separate once again with a period ${\xi}$$_{p}$,
given by \cite{desem_chu}
\begin{equation}
\xi _{p} =\frac{\pi \sinh (2q_{0} )\cosh (2q_{0} )}{2q_{0} +\sinh
(2q_{0} )} .
\end{equation}
It is important to note the interaction force drops exponentially
with the distance between solitons.

Soliton interactions for two solitons, three solitons, and a
random sequence of solitons have been studied numerically \cite
{pinto_agrawal}. It was found that in a given soliton stream with
a number of consecutive bits filled, it only the first and last
solitons of the consecutive bit stream are effected by the soliton
interaction because the interior solitons experience approximately
equal forces from each side.
\item \textbf{\textit{Third Order Dispersion}} \\
Dispersion in fibers is usually described by taking the lowest
order term in a group velocity dispersion (GVD) expansion. In most
cases, this term is dominant and the first order expansion is a
good approximation. However, near the zero point of dispersion,
the next term in the expansion, known as third order dispersion
can be of comparable magnitude to the lowest order term, and must
be considered \cite{agrawal}. This effect has attracted attention
recently because it can be a significant effect in a
dispersion-managed system, where the fiber alternates between
being dispersive and anomalously dispersive and the average
dispersion is slightly anomalously dispersive and very small
\cite{hizanidis}. In general, this effect is small unless the
soliton frequency is very near the zero of dispersion
(${\beta}$$_{2}$) for the fiber.
\item \textbf{\textit{Polarization mode cross coupling}} \\
Even in a single mode optical fiber, two polarization modes can be
maintained. Dispersion between these two modes is known as
Polarization Mode Dispersion (PMD.) The internal structure of a
soliton is very stable under the effects of PMD. However, the
different polarization modes most often have different linear and
non-linear indexes of refraction and therefore travel at different
group velocities. Over a long span of fiber, even a pulse that
starts initially in one polarization mode may be split into both
modes via PMD and therefore reach the end of the fiber at
different times. In addition, noise involved in PMD causes a
soliton timing jitter \cite{agrawal} \cite{mollenauer_gordon}.
\item \textbf{\textit{Temperature fluctuations}}
Temperature fluctuations can cause a substantial change in the
optical path of a pulse during storage in a loop. This effect can
be large compared to the delay induced by the gravitational
frequency shift and large compared to the other effects listed
above.
\end{itemize}

\subsection{Design issues and performance}
\label{sec-design-perf}

Careful design can help control the various sources of noise
listed above. The Raman self frequency shift and soliton-soliton
interactions can be limited by using relatively wide solitons
(\texttt{>}1 ps) at a relatively low bit rate (10 Gbit/s). Also,
soliton-soliton interactions can be reduced by the use of
synchronous amplitude modulation \cite{nazakawa} and perhaps
amplitude variation. In addition, one can reduce the effect of
soliton-soliton interactions by using the relative timing between
interior bits only \cite{pinto_agrawal}. Third order dispersion
can be made negligible by staying away from the zero of
dispersion. In fact, the signal in this model depends on the
solitons seeing a net dispersion as they transit the loop.
Polarization mode cross coupling can be reduced through the use of
polarization maintaining (PM) fiber in the loop. However, it would
increase the fiber cost substantially if a large fraction of the
loop were constructed with PM fiber. Some of the effect of
polarization mode cross coupling can be removed by making a
differential measurement. In this model it would be necessary to
place polarization controllers in strategic locations throughout
the circuit. In addition, it might be possible to use polarization
specific filtering and detection or make use of trapped or vector
solitons (where the two polarization modes have the same group
velocity)\cite{agrawal}.  The timing clock signal for the
synchronous modulation can be extracted from the transmitted
solitons themselves, following Nazakawa et. al. \cite{nazakawa}

Even taking into consideration proper design, there will exist
residual relative timing jitter and shifts from the noise sources
above. We can place a reasonable limit on the size of the random
jitter by noting the results of previous bit storage experiments.
Nazakawa et al. established long-term storage of 24 ps wide
solitons at 10 Gbit/s \cite{nazakawa}. Numerous other groups have
transmitted soliton bit patterns for shorter distances at higher
(20-160 Gbit/s) rates \cite{stor_loops}. Liao and Agrawal have
shown through numerical simulations that it should be possible to
limit the jitter to approximately 1 ps at a rate of 40 Gbit/s
\cite{liao_agrawal}. Based on these results, it seems reasonable
to assume the random jitter can be controlled to better than the
10 ps level. This corresponds to an uncertainty in the pulse
position at detection of approximately 2 mm, ignoring thermal
effects for the moment.

Now, let us return to our experimental model and consider only
noise that can cause random jitter. A 10 Gbit/s soliton beam
propagating in a fiber with D=+15 ps/nm-km ($\beta_{2}=-19
ps^{2}/km$) is split into two streams. One travels vertically a
distance h and is injected into a fiber optic storage ring, while
the other is injected into a similar fiber optic storage ring at
the height of the original soliton pulse source. There will be a
gravitationally-induced shift in the group velocity of the
solitons that move vertically, according to equation (9). If the
storage rings are constructed of similar fiber (assume D=+15
ps/nm-km, SMF-28 fiber), the average, relative group velocity
difference between the streams of solitons traveling in the two
storage rings will remain. Over a storage time T, the light will
travel a distance d in the storage loops and the group velocity
difference will induce a relative shift in the length traveled
along the two paths (in meters) of
\begin{equation}
\delta d=1\times 10^{-10} hT .
\end{equation}

There will be
\begin{equation}
N=\frac{Ln\left( rate\right) }{c} =50L
\end{equation}
solitons in each storage ring of length L, assuming the rings are
filled. The statistical average error on the relative position of
solitons in one ring as compared to the other when the beams are
recombined is
\begin{equation}
\delta s=\frac{0.002}{\sqrt{N} }  meters.
\end{equation}

\begin{figure}[b]
\centerline{\epsfig{file=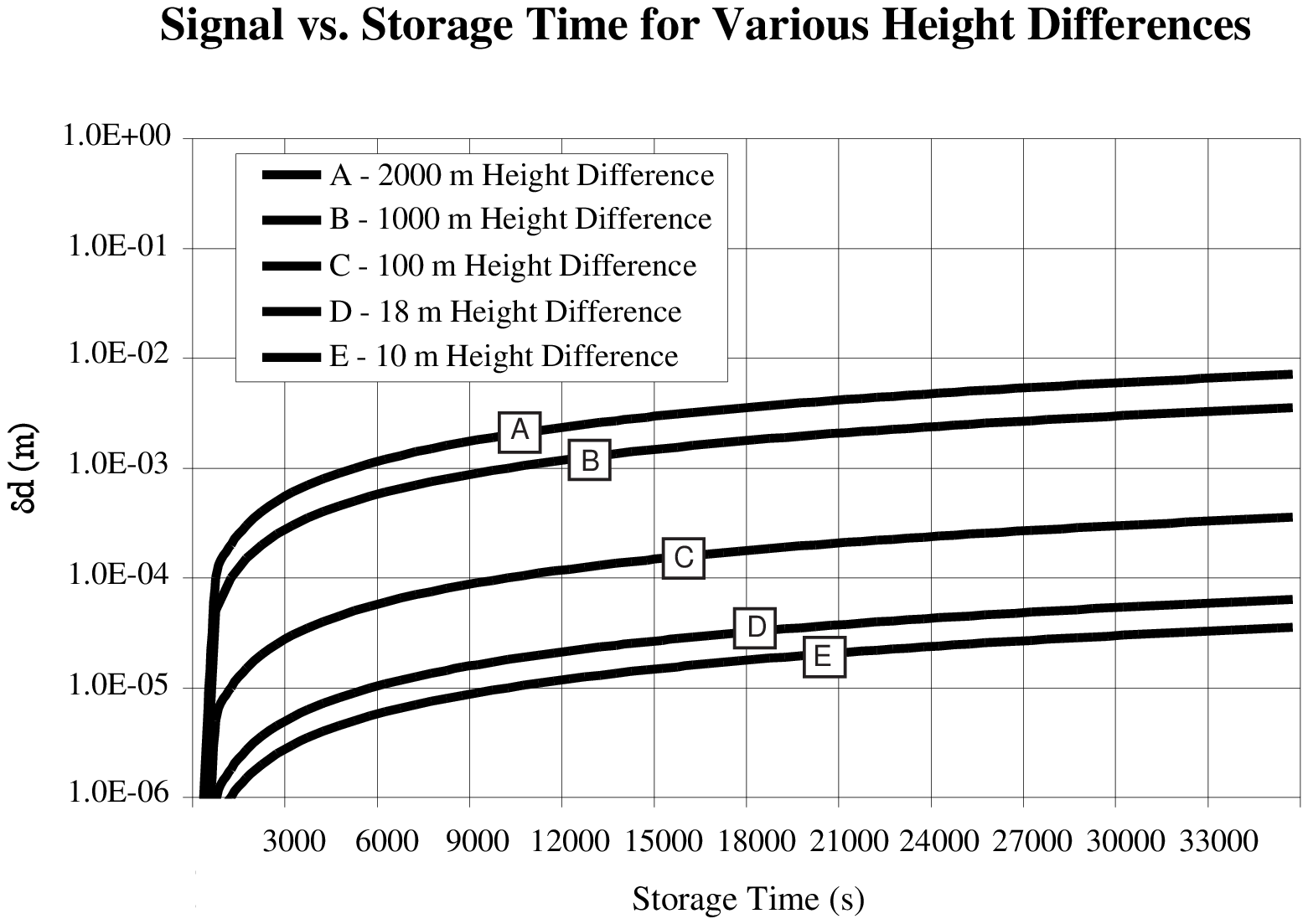,width=9cm} }
   \caption {Difference in travel distance between pulses in
vertically separated storage loops as a function of storage time.}
   \label{signal-vs-storagetime}
\end{figure}

\begin{figure}[b]
\centerline{ \epsfig{file=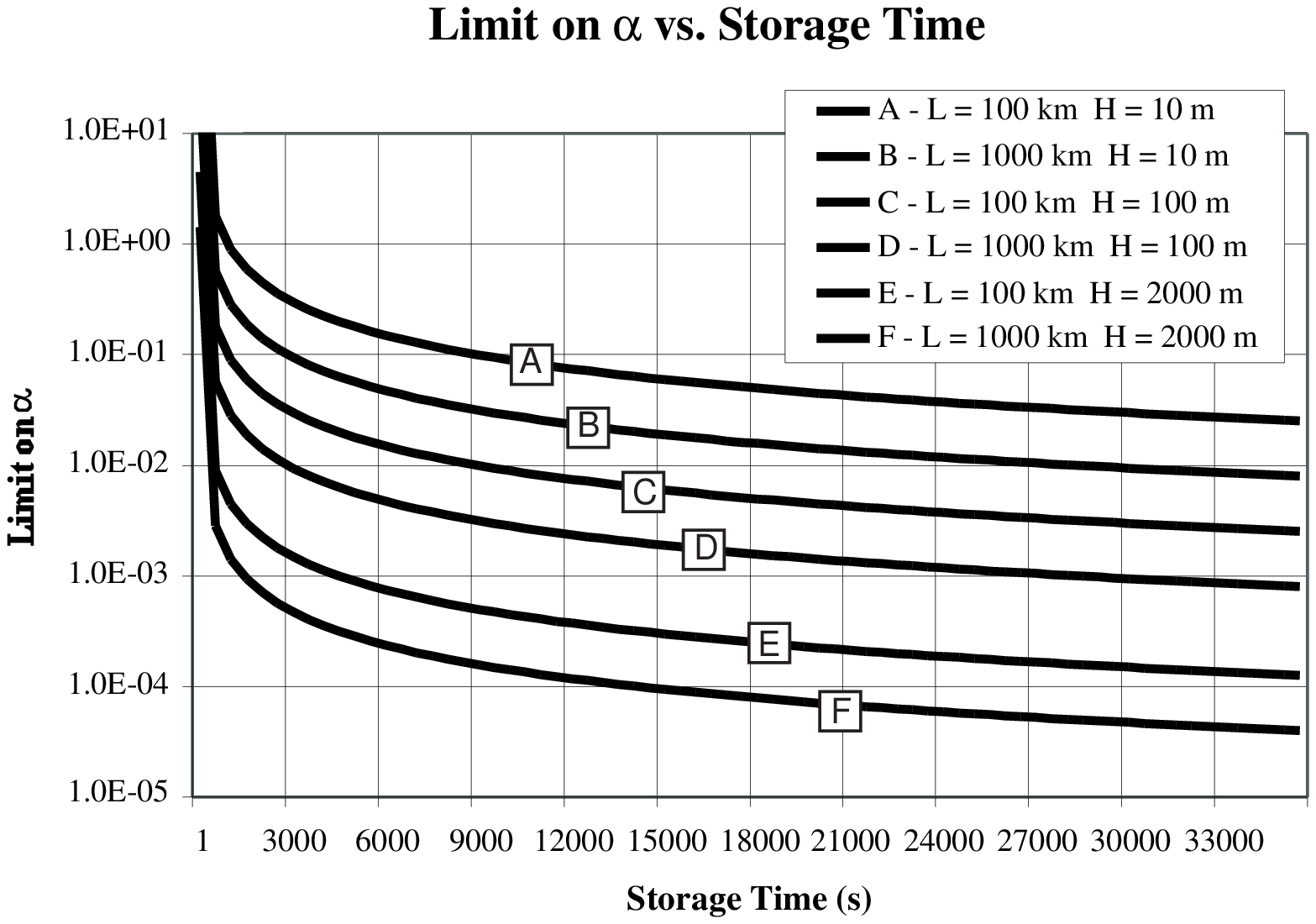,width=9cm} }
\caption{Sensitivity to ${\alpha}$ as a function of storage time
for various parameters of the two storage loop scheme.}
   \label{limit-alpha-earth}
\end{figure}

Figure~\ref{signal-vs-storagetime} shows ${\delta}$d as a function
of vertical loop separation d, and storage time T. The sensitivity
to ${\alpha}$, given by ${\delta}$s/${\delta}$d, is plotted in
Figure~\ref{limit-alpha-earth}. This figure shows an interesting
limit on ${\alpha}$ can be achieved under circumstances where only
random jitter noise effects are considered.

Unfortunately, unidirectional soliton timing shifts within each
ring are unavoidable and quite large. However, so long as these
shifts are reproducible and consistent for a given ring, it should
be possible to extract the gravitationally-induced timing shift
using differential measurements between different loop heights.
Our conceptual design creates a stable environment for the storage
loops in hopes of achieving this reproducibility.

\begin{figure}[b]
\centerline{
\epsfig{file=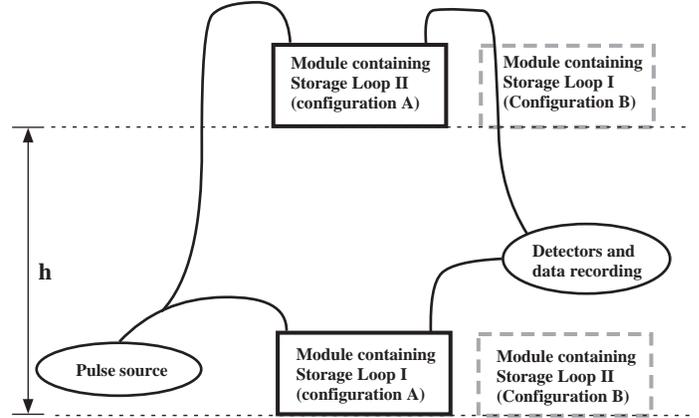,width=9cm} }
\vspace{10pt}
   \caption {The basic concept of the two storage loop differential
measurement.
   \label{basic-soliton}}
\end{figure}

Figure~\ref{basic-soliton} illustrates the basic concept of making
differential measurements in this experimental model. The fiber
storage rings are constructed as ``modules'' that can be moved
carefully from place to place without disturbing the active
elements of the storage loop or the environment that surrounds
them. Fiber leads to and from each module are detachable, so the
relative spatial configuration of the two modules can be changed
in order to make differential measurements. Between separate
differential measurements there will be relative changes in the
input and output fibers. However, the light spends the vast
majority of the storage time inside the loop. The input and output
fibers constitute an extremely small fraction of the total path
length, and will have a correspondingly small effect on the pulse
timing.

\begin{figure}[b]
\centerline{ \epsfig{file=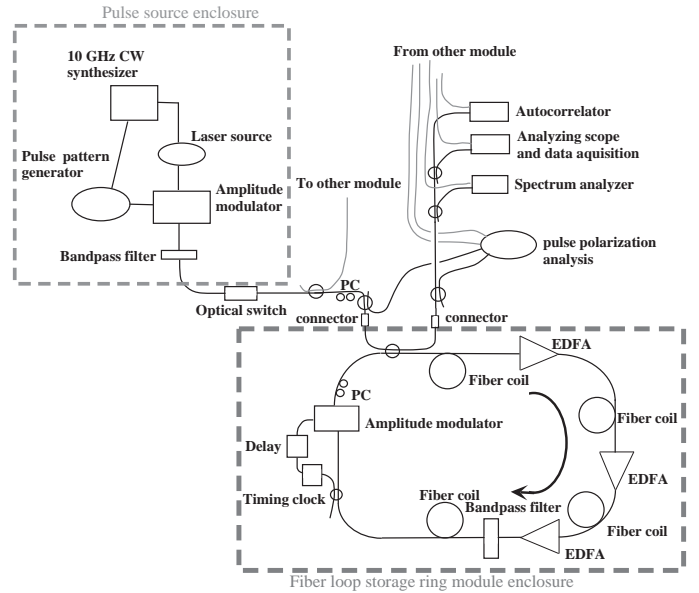,width=9cm} }
   \caption {Conceptual design of the optical circuit.}
   \label{conceptual-soliton}
\end{figure}

Figure~\ref{conceptual-soliton} shows the basic layout of the
experimental optical circuit in more detail. The source and both
storage rings (and detectors, if required) are in stable
environments. The ring modules are constructed to identical, and
can be disconnected and moved to different heights, as needed.

Systematic effects, such as those arising from small differences
between the modules/storage rings, can be taken out with
differential measurements. Each pass around the ring, a small
fraction of the power is picked off so that the relative timing in
the two rings can be compared.

\begin{figure}[b]
\centerline{ \epsfig{file=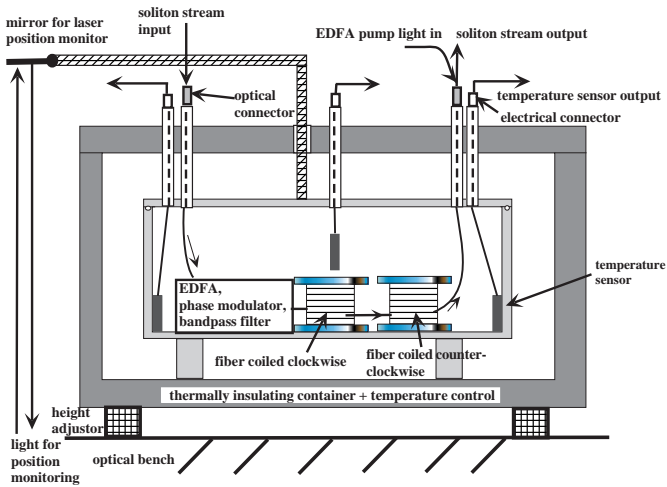,width=9cm} }
\vspace{10pt}
   \caption {Conceptual design of a fiber optic storage ring module.}
   \label{module}
\end{figure}

Figure~\ref{module} shows one concept of a fiber optic storage
ring module. It consists of the basic storage ring components
(fiber, bandpass filter, EDFAs, and amplitude modulator) in a
thermally isolated compartment that can be moved as needed. The
fiber is coiled on spools that have low thermal expansion
coefficients. The coils are arranged so the net area subtended by
the loop is very small. This will reduce any effects due to the
rotation of the earth during the measurement.

The module is placed in a carefully controlled, stable thermal
environment. The temperature inside the module is carefully
monitored. If necessary, this information can be used for a
feedback system to keep the temperature stable over time. The
absolute accuracy of the temperature sensors is not critical.
However, they need to give reliable relative measurements over
time.

Similarly, the pressure in each module is monitored and adjusted
slightly as needed.  The relative pressure between the inside of
the module and the outside is relatively small in order to
minimize difficulties in module enclosure construction.  We plan
to seal the modules while at the same height and pressure and then
transport them to different vertical positions while maintaining a
constant internal pressure.

Unfortunately, relatively small fluctuations in the ambient
environment of the loops can cause large effects over such long
storage times. For example, assuming a coefficient of thermal
expansion for the fiber of approximately 10$^{-7}$/$^{\circ}$C and
a relative temperature difference between two storage loops of
$\delta$t $^{\circ}$C leads to a fractional optical path length
difference of 10$^{-7}\delta$t. This compares to the
gravitationally induced effective fractional path length
difference of approximately 10$^{-19}$h, for a loop height
separation of h meters. It is impractical to assume the
temperature in the modules can be controlled to high enough
accuracy to make these numbers even remotely comparable.  Rather,
it will be necessary to calibrate out effects due to temperature
and pressure variations.  One way of doing this is to use
wavelength division multiplexing in the soliton bit stream.  For
example, suppose the soliton bit stream contains pulses centered
at the zero-dispersion frequency of the fiber as well as pulses in
the dispersive region.  The relative pulse timing between the
zero-dispersion-centered solitons gives a measure of the relative
environmental fluctuations since the gravitational effect is
nonexistent at the zero dispersion frequency.  The relative pulse
timing between the pulses in the dispersive region will contain
the environmental fluctuations and the gravitational effect.  In
principle, the gravitational effect can be isolated by subtracting
out the relative timing difference of the zero-dispersion-centered
solitons.  The ultimate success of this experiment hinges, in
part, on how well this can be done.

All wires and fibers leading into the module are detachable,
so the modules can be exchanged with ease, without disturbing
the active elements of the loop. The items inside the module
are secured so that minor movements of the module leave no net
change in the positions or stresses inside the module.

A careful survey is done to locate the modules with respect to
each other and to positions where measurements of the local
gravitational field strength are done using a reasonably accurate
gravimeter (good to 1 part in 10$^{6}$ or 10$^{7}$).

Each module rests on a stable optical table. The relative position
of the two modules is monitored during a store via an interferometric
system (or repeated optical surveys if such a system is simpler
and more cost effective). Significant changes, perhaps due to
diurnal variation in the building, can be taken out via small,
remote control, height adjusters located underneath one of the
modules. If necessary, these adjustments could be made automatically
with feedback from the interferometric system. Accuracy on the
order of 200 microns should be achievable and sufficient.

The change in the group velocity, and therefore the signal size,
is directly proportional to the group velocity dispersion in the
fiber loop. It is necessary to measure this dispersion as
precisely as possible so it does not limit the measurement
precision on ${\alpha}$. Measurements of fiber chromatic
dispersion to ${\pm}$0.02 ps/nm-km have been reported
\cite{christensen}. Fujise \textit{et al.} use a
wavelength-division-multiplexing (WDM) phase shift method which
compares the relative phase shift between modulated light of
different wavelengths passing through the fiber simultaneously.
They achieve a measurement stability of ${\pm}$0.02 ps/nm-km over
a ten-hour period. We will need to improve on this precision, or
develop the ability to correct for changes over time using this
(or a similar) technique in situ as a monitor. Another method we
are considering is to send separate soliton pulses with different
carrier frequencies through the loop and look at the relative
timing between the pulses before and after the loop. The stability
of our dispersion measurement is intimately tied in with the
stability of the overall timing at the end of a store and our
ability to extract the gravitational effect using differential
measurements.

\subsection{Potential for a flight project} \label{sec-fight-expt}

Since the signal size of the soliton ring model depends linearly
on the height difference of the two loops in a given gravitational
field, it is desirable to make this quantity as large as possible.
To this end, it is interesting to explore the potential gains and
difficulties that might be encountered performing the soliton
storage ring experiment in orbit on the space shuttle, or the
International Space Station. In such an experiment, one storage
loop would remain on the shuttle/station, while the other loop
would be transported to an orbit with a different radius or
height, but the same angular velocity. In this scheme, loop height
differences of tens of kilometers or more could be achieved.

\begin{figure}[b]
\centerline{ \epsfig{file=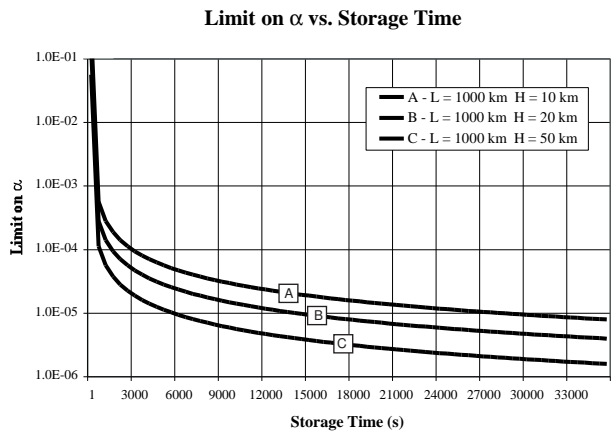,width=9cm} }
   \caption {Potential sensitivity to ${\alpha}$ for different
loop separations in a space-based experiment. }
   \label{limit-alpha-space}
\end{figure}

The acceleration due to gravity in orbit will differ from that on
the surface of the earth by about 10\%. (The value of g is around
8.94 m/s$^{2}$ at an altitude of 160 nautical miles, which is the
altitude of STS-101.) This comes in the signal with a linear
dependence and must be taken into account for an exact
calculation. However, for simplicity, we will ignore this
difference, along with the change in g as a function of height
along the loop. In practice, these (O(10\%) effects) must be
measured and taken into account. With this simplification,
equation (14) is still valid. Figure~\ref{limit-alpha-space} shows
the potential sensitivity to ${\alpha}$ for heights of 10, 20, and
50 km, respectively.

Conceptually, one module remains on the shuttle (or space station),
while the other is ensconced in a small, unmanned vehicle that
contains environmental control, power for EDFA pump lasers and
thermal control, environmental monitoring, devices capable of
relating the exact position of the one module with respect to
the other, and minor propulsion and navigational capability for
stability (and possible maneuvering, though perhaps this is better
done by the shuttle). The two modules must be tethered together
by the fibers that carry the soliton stream to and from the second
module. The fibers should be surrounded by a thin, flexible covering
that evens out the azimuthal asymmetry in the thermal environment
and provides some degree of stress relief.

Conceivably, the solitons could be generated separately on the
two modules with the timing communicated via light pulses sent
back and forth. This has the advantage of removing the need for
the fiber tether and allowing the possibility of a greater height
difference, and perhaps a sensitivity to higher-order terms.
Unfortunately, this raises a host of other issues, such as relative
differences in the loop soliton sources and gravitational effects
on the clock rate of electronics that communicates between the
modules.

It is also important to note that the primary elements of this
experiment are small, light, and mechanically robust. In addition,
the experiment requires a relatively modest amount of power.
As such, it could be flown on a variety of platforms.

This experimental model is quite novel. It would have very
different systematic errors than the clock-in-space experiments
that are underway.

\section{Experimental model of vertical fiber optic Sagnac
interferometer with varying dispersion} \label{sec-expt-sagnac}

Coherent light traveling from a source is split into two beams
that travel in opposite directions around a loop or coil of
optical fiber. After traversing the coil, the two beams of light
are recombined and the phase difference between the clockwise (CW)
and counter-clockwise (CCW) beams is measured. This technique is
used widely to make high precision fiber optic gyroscopes (FOG)
that make use of the fact that a phase difference between the two
paths is induced by a rotation of the coil \cite{FOGs}. In our
model, the coil lies in a vertical plane. Unlike the FOG, the
fiber dispersion must vary asymmetrically around the circumference
of the loop in order yield a net phase shift due to the interplay
of dispersion and the gravitational frequency shift. At the top
and bottom of each loop in the coil are splices to join the fibers
of differing dispersion. The pattern of varying dispersion is such
that the CW (CCW) beam always travels upward through fiber with
positive dispersion and downward through fiber with negative or
anomalous dispersion. The situation is reversed for the CCW (CW)
beam.

As the CW and CCW beams propagate through the coil, a phase
difference between the beams develops with time due to the
gravitational frequency shift. This phase difference is given by
\begin{equation}
 \delta \varphi =\frac{2\pi Ngh^{2} \nu ^{2}
}{nc^{3} } \left[ \left( \frac{dn}{d\nu } \right) _{L} -\left(
\frac{dn}{d\nu } \right) _{R} \right]
\end{equation}
for beams of initial frequency ${\nu}$, traveling N turns around a
loop of height h in a uniform gravitational field with strength g.
The dispersion on one side of the fiber loop is
(dn/d${\nu}$)$_{L}$ while that on the other side is
(dn/d${\nu}$)$_{R}$.

At 1550 nm, for an interferometer constructed of fibers with
dispersion +15 ps/nm-km (=3.7x10$^{-17}$ s) and \\ -90 ps/nm-km
(=-2.2x10$^{-16}$ s), respectively, and setting n=1.5, equation
(26) becomes
\begin{equation}
\delta \varphi =\left( 1.6x10^{-11} \right) Nh^{2}.
\end{equation}

\begin{figure}[b]
\centerline{ \epsfig{file=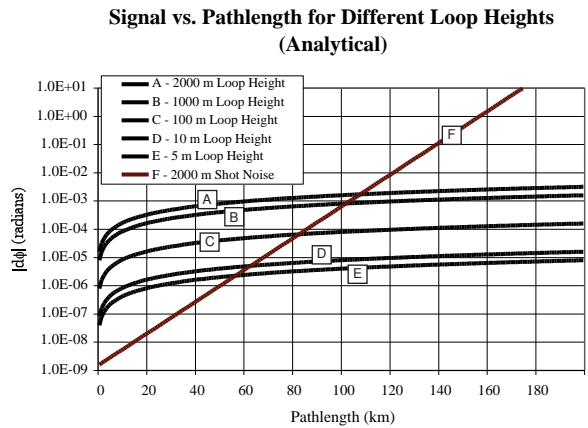, width=9cm} }
   \caption{Signal size versus path length for various loop heights
along with the shot noise for a 2000 meter loop. (Results from an
analytical calculation.)}
   \label{sagnac signal ana}
\end{figure}

Figure~\ref{sagnac signal ana} shows the size of the phase shift
as a function of total path length and loop height. Although many
sources of noise need to be considered, the measurement precision
is limited ultimately by light loss in the fiber and at the
splices.

In order to estimate, roughly, the potential sensitivity of such
an experiment, it is necessary to fold in the shot noise. The
photon counting noise in the phase can be estimated by the
Heisenberg uncertainty principle as
\begin{equation}
\left( \delta \varphi \right) _{shot} \approx
\frac{1}{\sqrt{N_{photons} } }.
\end{equation}

For a 10 mW laser source operating at 1550 nm, this estimate
becomes
\begin{equation}
\left( \delta \varphi \right) _{shot} =\frac{1}{\sqrt{\left(
5x10^{17} \right) te^{-\alpha d} } },
\end{equation}
where t is the signal integration time (in s), ${\alpha}$ is the
total fiber light loss (in dB/km), and d is the total distance
traveled by the light (in km). Assuming a fiber light loss of 0.5
dB/km and a loss at each splice of 0.5 dB yields the shot noise
limit shown in Figure~\ref{sagnac signal ana}.

The results in figure~\ref{sagnac signal ana} are from analytical
calculations. We have also used the split step Fourier method to
solve the non-linear Schrodinger equation for this model
numerically \cite{mccoy}. The results, shown in Figure~\ref{sagnac
signal num}, are very similar to those obtained by analytical
calculations. The difference in the shot noise curve comes about
because the analytical model treats the light loss at splices as a
continuous process, while the numerical model incorporates
discrete losses at the splices and should be more accurate.

Things besides shot noise can limit the sensitivity of this
experiment. Source relative intensity noise, thermal fluctuations,
changing mechanical stresses, polarization mode cross coupling,
and coherent backscattering at the splices can lead to
nonreciprocal phase shifts. In addition, electronics noise can be
an issue. Without constructing a very detailed model and/or
building prototypes, it is difficult to estimate the level of
these sources of noise in this experimental model. One useful
point of comparison is the fiber optic gyro, mentioned earlier.
The FOG is a continuous wave fiber Sagnac interferometer that is
vulnerable to many of the same sources of noise. Very high
sensitivity FOGs are sensitive to a rotation-induced phase change
of 1x10$^{-9}$ radians. A sensitivity of one ${\mu}$rad is
considered difficult, but achievable \cite{FOGs}
\cite{fibersense}.

\begin{figure}[b]
\centerline{ \epsfig{file=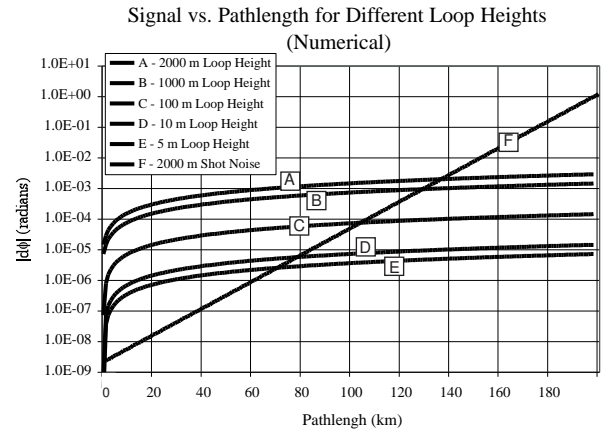, width=9cm} }
   \caption{Signal size versus path length for various loop heights
along with the shot noise for a 2000 meter loop. (Results from a
numerical calculation.)}
   \label{sagnac signal num}
\end{figure}

\begin{figure}[b]
\centerline{ \epsfig{file=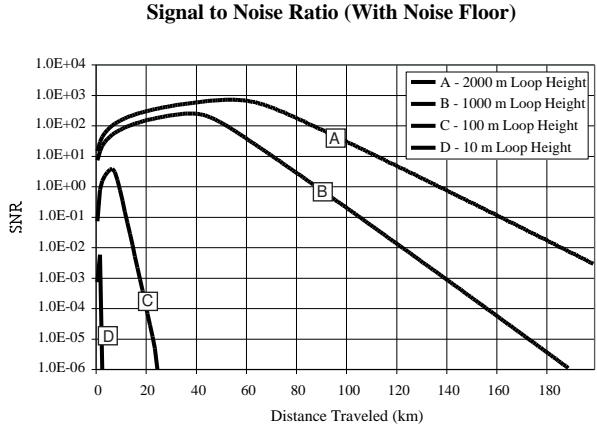, width=9cm} }
   \caption{Signal to noise ratio, assuming a noise floor at 1 ${\mu}$rad
and shot noise.}
   \label{snr w noisefloor}
\end{figure}

Figure~\ref{snr w noisefloor} shows the signal-to-noise ratio for
this model with varying loop height, shot noise and a 1 ${\mu}$rad
noise floor. Without using squeezed light, the shot noise provides
a hard cutoff in the path length for a given loop height. An
optimistic estimate of what could be done using squeezed light is
shown in Figure~\ref{snr squeezed light}. For this curve, the shot
noise is reduced by a factor of 100 \cite{boyd}.

In our view, Figures~\ref{snr w noisefloor} and ~\ref{snr squeezed
light} represent an optimistic assessment of what might be
achievable with current technology. Our model differs from a FOG
in some very significant ways. FOG designers struggle to maintain
reciprocity between the counter propagating beams in order to
achieve the high sensitivity of the best instruments.
Non-reciprocal dispersion is a necessary component of this model.
In addition, FOGs are wound about titanium cores in a carefully
maintained and monitored environment. In our model, the loop
necessarily extends a large vertical distance, which would
complicate attempts at noise reduction.

Another weakness of this model is that it would be very difficult
to make a differential measurement, which would be helpful for
extracting the signal. It is hard to imagine being able to rotate
the device or change the height without causing arbitrary phase
changes between the counter-propagating beams. The growth of the
phase shift with time, compared in different orientations, would
be the best way to pull out the signal.  This is another place
where our model differs significantly from the FOG. FOGs work with
continuous wave light and our model requires long pulses. 0

To summarize our work on this model, it is conceivable that a high
precision measurement of the gravitational redshift of light could
be made using a vertical fiber Sagnac interferometer with varying
dispersion. However, the technical difficulties are challenging
and further modeling and prototyping would be useful in evaluating
the eventual sensitivity that might be achieved.

\begin{figure}[b]
\centerline{ \epsfig{file=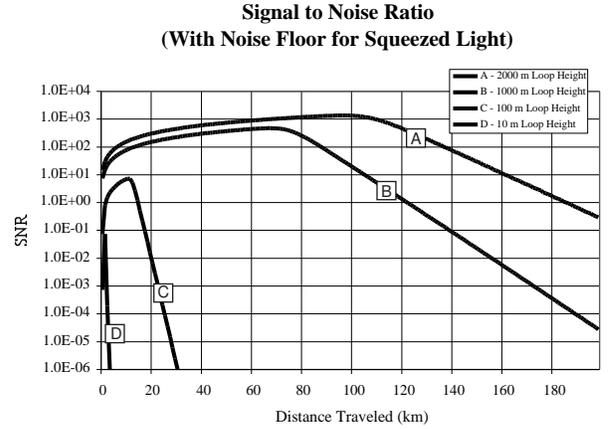, width=9cm} }
\caption{Signal to noise ratio for the Sagnac interferometer model
with various loop heights and a 1 mrad noise floor.  The shot
noise included in this calculation is reduced by a factor of 100,
relative to Figure~\ref{snr w noisefloor}, as an optimistic
assessment of the effect of using squeezed light.} \label{snr
squeezed light}
\end{figure}

\section{CONCLUSIONS}
\label{sec-conclude}

The interplay of optical dispersion and the frequency shift of
light due to a changing gravitational potential can be used to
test the equivalence principle for gravitation by converting the
frequency shift into a phase shift or time delay.  It is possible
an experiment using solitons in optical fiber storage loops could
be performed that uses this effect to make a high precision
measurement of the gravitational redshift of light.

\section{Acknowledgements}

We are indebted to the following people for useful discussions:
Govind Agrawal, Robert Boyd, Jeffrey Clark, Turan Erdogan, Alan
Evans, John Heebner, Daniel Kerr, Jeff Korn, Taras Lakoba, Adrian
Melissinos, John Moores, Michael Perlmutter, Sarada Rajeev, and
Ram Yahalom. This work was supported by the University of
Rochester.

\end{document}